# An Integrated Approach for Extraction of Objects from XML and Transformation to Heterogeneous Object Oriented Databases

Uzair Ahmad, Mohammad Waseem Hassan, Arshad Ali
*National University of Science & Technology (NUST) Tamiz Uddin Road, P.O.Box 297, Rawalpindi, Pakistan*

Richard McClatchey
*Centre for Complex Cooperative Systems (CCCS) UWE, Bristol BS16 1QY UK*

Ian Willers
*European Organization for Nuclear Research (CERN) 1211, Geneva, Switzerland*

Abstract: CERN's (European Organization for Nuclear Research) WISDOM project uses XML for the replication of data between different data repositories in a heterogeneous operating system environment. For exchanging data from Web-resident databases, the data needs to be transformed into XML and back to the database format. Many different approaches are employed to do this transformation. This paper addresses issues that make this job more efficient and robust than existing approaches. It incorporates the World Wide Web Consortium (W3C) XML Schema specification in the database-XML relationship. Incorporation of the XML Schema exhibits significant improvements in XML content usage and reduces the limitations of DTD-based database XML services. Secondly the paper explores the possibility of database independent transformation of data between XML and different databases. It proposes a standard XML format that every serialized object should follow. This makes it possible to use objects of heterogeneous database seamlessly using XML.

## 1  Introduction

XML has rapidly emerged as the standard for exchanging business data on the World Wide Web. This makes the ability to transform database data to and from XML an increasingly important issue to resolve. XML's nested and self-describing structure provides a simple yet flexible means for applications to exchange data. If XML is to fulfill its potential, some mechanism is required to transform data between databases and XML. The transformation of database data into XML and back has therefore become an important requirement of modern database management systems.

Database and XML related research is regarded as the most promising and challenging direction for the community of database researchers (Bernstein 1998, Ceri 2000). Many research communities are involved in exploiting the platform-independent and self-describing nature of XML for database applications some examples being (Michael 2000, Bhatti 2000, Jayavel 2000).

There are two inter-related issues that this paper addresses. The first issue is the incorporation of the W3C (World Wide Web Consortium) XML Schema specification in the XML-Database relationship. Previously, the W3C Document Type Descriptors (DTDs) were essentially used as schemas for XML documents. These DTDs were developed for domains as diverse as electronic commerce (Down 2000) and scientific domains such as that at CERN (Bhatti 2000, Afaq 1999). As technology advanced, updating of existing solutions became inevitable. Similarly, the finalization of XML Schema specifications has created a need to exploit its potential for the benefit of XML services for databases. This paper specifies a new approach that incorporates recent XML advances to help improve existing facilities of data transformation from databases into XML and back. Object databases are taken as the target database platform for this work. The remainder of this paper refers to our system "Transformation of data between XML and object databases" as **TransODB**.

The second issue addressed in this paper surrounds the offering of database-XML services. Currently services offered by different database vendors are not standardized and these XML services are often database specific. TransODB explores the XML potential to represent objects of heterogeneous object databases in a standardized way. This work proposes a database independent solution of transferring objects in different databases using XML. This is highly desirable in heterogeneous database environments.

Making data resident in different databases understandable and shareable by multiple different clients

(Ceri 2000, Jayavel 2000) is a challenging problem. In an heterogeneous database environment, such as that prevalent at CERN, where clients need information that reside in different physical data repositories the problem becomes even more difficult to solve. In these environments typically a user of one particular database may need to get certain information from a foreign database e.g. Objectivity/DB and store that information locally on some other database e.g. ORACLE9. This requires a middleware schema that can describe data residing in different databases. This can be achieved using the TransODB concept of data transformation into a specialized XML form. TransODB implements this idea that incorporates specialized usage of W3C XML Schemas.

## 2 Related work

A number of communities have been interested in exploiting XML as the business data sharing standard in their frameworks. There is much work that can be regarded as relevant, if not fully related, to TransODB in one respect or the other. Here is a brief description of the work that was found to be helpful in building a conceptual makeup of the TransODB database.

### 2.1 WISDOM

WISDOM (Wide-area database Independent Serialization of Distributed Objects for Migration) components replicate data in distributed environments (Bhatti 2000, Afaq 1999). These components were developed when, for XML document structure, DTDs had been used. WISDOM components are deficient in certain important features. XML has improved significantly since it was used in the WISDOM setup. Since they are developed prior to the W3C XML finalization they lack the real strength of XML, especially the XML Schema. Secondly, platform neutrality (an important motive for switching to Java) is not present in the previous solution. On the database end, Objectivity offers more flexibility and power to Java applications than C++ applications. In the absence of the technologies much work is necessary in developing WISDOM components. Incorporating these new technologies has significantly simplified the problem.

### 2.2 Objectivity/DB XML Interface

Keeping in mind the status of XML as the leading standard for the transfer of information between business systems, Objectivity/DB also provides an XML interface for object to XML transformation. This interfaces provides a link between objects stored in Objectivity/DB and any given XML representation (Objectivity Inc). The Objectivity XML Interface Tool supports the export of objects from the latest version of Objectivity/DB in XML format and the import of XML formatted data into Objectivity DB. Objectivity aims to provide open source for the XML Interface Tool free of charge to its customer base. These interfaces incorporate two modules, ooXMLdump and ooXMLload.

The ooXMLdump utility transforms the schema and objects (data) into "well-formed" XML documents. The ooXMLload utility transforms XML-ized objects into live database objects. These utilities incorporate various degrees of compression to improve performance. New interfaces are also expected to show an improvement in performance over the older tools. Currently these utilities use DTD-based XML files (Micram AG). Objectivity is going to incorporate the W3C XML Schema into its XML interfaces in its next release. Moreover Objectivity provided utilities work only for Objectivity objects. TransODB makes this transformation database independent.

## 3 TransODB Design features

As described earlier the nature of the problem comprises three internally different domains. At each domain level a number of competing options have emerged. Usage of XML for database schema description remains an important concern for TransODB. For XML schema description there are six candidate languages [9]. TransODB incorporates the latest W3C recommendations of the XML Schema for describing XML documents.

### 3.1 XML Schemas

The impact of the XML Schema on the database XML relationship is discussed in this section. In the TransODB domain the XML-ized database information is in two parts. The first part represents the DB Schema of the object database and the second part represents the object data residing in the database according to this DB Schema.

#### 3.1.1 DTD based approach

The generally adopted practice of describing the structure of XML documents involves the usage of DTD. Currently the WISDOM architecture follows the same approach for representing the object database in XML. Figure 1 shows how WISDOM maps Object database contents into XML.

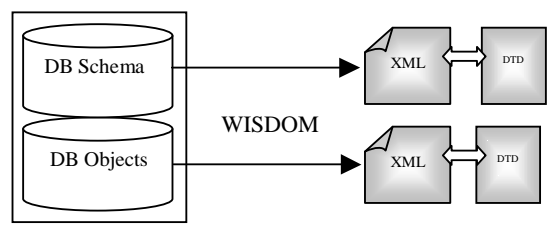

Figure 1. XML-ized database view – common approach

#### 3.1.2 TransODB approach

The TransODB approach involves only two XML files for the whole database representation. Figure 2 shows the simple and straightforward replica of unnecessarily complex XML content that was used in DTD based solutions.

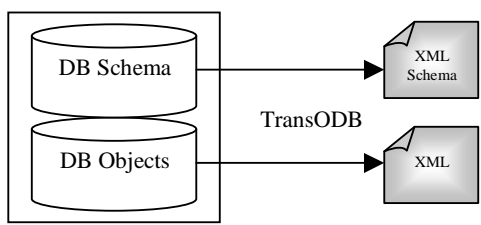

Figure 2. XML-ized database view – TransODB approach

For describing the schema of Object database it uses only one file of XML Schema. And for representing objects (data) of the database, it uses one XML file. This reduces the number of files to represent Object Database perfectly, and takes less XML content than the DTD based approach.

## 4   TransODB Architecture

This section describes TransODB architecture. It has two modules. Firstly it converts the existing object database into XML and then, in a second module, it recreate the objects from XML.

### 4.1   Converting database to XML

The TransODB concept proposed by this paper expects specialized XML input. Certainly the way we remake the database must determine how we should convert the database to that particular XML format.

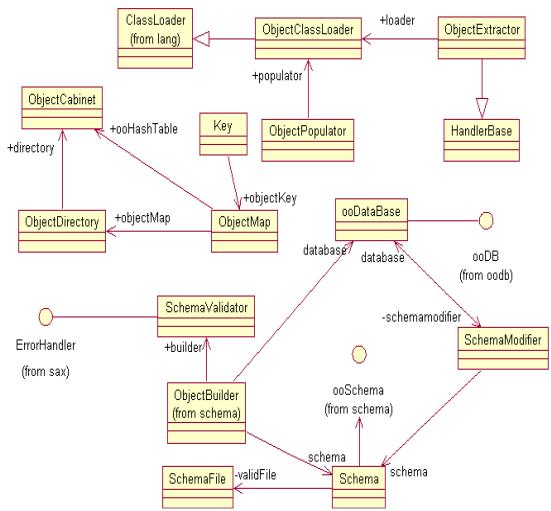

Figure 3. TransODB class Model

TransODB also proposes specialized XML generation methodology.    Figure 3 shows the UML class diagram of the TransODB XML to Database Conversion module. TransODB adopts an XML Schema oriented methodology. Despite using XML for describing object database schema, it incorporates XML Schema files.

This paper advocates the importance of a lightweight XML-ized database for substantial improvement in data migration. By saying lightweight we mean the minimum possible amount XML data to represent a database perfectly.

### 4.2   XML to Database Conversion

The TransODB design is based on a comparative study of technologies and an investigation of object databases and XMLThe TransODB design uses the latest parsing methodologies and database interfaces. A comprehensive description of the TransODB design is given below.

#### 4.2.1   DB Schema Builder

As discussed earlier TransODB must cater for both the design of the database (DB Schema) and data of the database (DB Objects). This module has the powerful feature of converting the XML schema files into corresponding Java classes. The DB Schema builder uses the SAX 2.0 parser for generating events of the XML Schema documents. After creating respective class objects of target XML Schema documents it starts generating Java code. The generated code follows the Java coding standards defined by Sun Microsystems. This module is responsible for extracting the DB Schema out of the XML Schema file. It incorporates Java reflection, the SAX parser and an event handler for the parser events, and a set of Java classes that enables SAX generated events to take shape into a Java class.

#### 4.2.2   Object Builder

The Object Builder module is responsible for the second major task of remaking objects and in essence is the heart of TransODB. It faces the challenge of remaking highly complex object networks residing in XML documents. Furthermore the documents are of significant volume. The Object Builder not only needs to read object semantics perfectly but to rebuild it in a database in the same form as in the originating database. The Object Builder consists of a set of sophisticated objects that perform the complex job of object rebuilding in the object database.

#### 4.2.3   DB-Handler

The object database recognizes Java classes and prepares its schema from them. The DB-Handler creates a class model from the Java classes extracted by the DB Schema Builder module. The second task of rebuilding objects is accomplished by an interaction with the Object Builder. The DB-handler uses object database given APIs for storing objects in the database.

# 5     Results of prototype

This section presents results of the prototype implementation of TransODB in comparison with performance readings of WISDOM and Objectivity/DB XML interface.

## 5.1     TransODB XML-ized Database

It is quite possible to describe all types of class constructs in the W3C XML Schema. Even if it is not meant for describing class models it is in every respect a faithful replica to the existing approach. Earlier, the DB Schema was described through XML defined by another DTD document that increases the complexity and workload unnecessarily. Figure 4 shows a comparison of TransODB XML-ized database file size with WISDOM file size for representing the same data

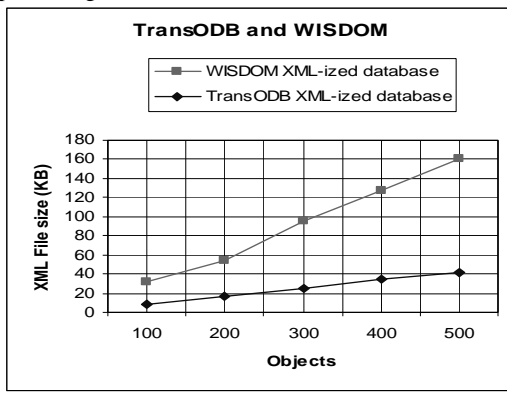

Figure 4: TransODB efficiency in terms of XML file size

## 5.2     TransODB and ooXMLload

Prototype tests have proved remarkably beneficial in terms of main memory usage.

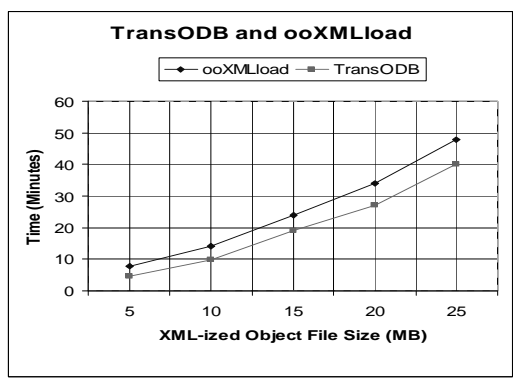

Figure 5: TransODB Performance readings

ing the objects does not unnecessarily load system resources and it allows other processes to run without any performance degradation. Figure 5 shows performance readings of TransODB in terms of time consumed for different file sizes. The TransODB proposed XML Schema oriented database has appreciably decreased the XML file size.

# 6     Conclusion

Transforming data into XML has become a common practice in the e-business world. Currently data to XML transformation tools are database and operating system dependant. This limits their usability in a heterogeneous environment like CERN. TransODB overcomes these limitations by incorporating the latest W3C XML Schema specifications and Java into the database-XML relationship. In the WISDOM project utilities, the XML-ized database contents comprise database storage details, which make the volume of the resulting XML documents many times greater than proposed by this paper. In TransODB, object-oriented database centric classes do not get XML-ized, this results in a significant decrease in the TransODB XML-ized database volume. Thus the specialized nature of XML-ized databases makes it possible to transfer objects of one database such as Objectivity into another database e.g. ObjectStore. In contrast, other alternative approaches can only handle objects of a particular database. Since TransODB has reduced the XML content to be processed both in number of objects and their size, a remarkable impact on TransODB scalability and performance has been achieved.